\documentclass[prl,floatfix,showpacs,preprint]{revtex4}
\usepackage{graphicx}
\usepackage{subfigure}
\usepackage{stdclsdv} \usepackage{array} \usepackage{amsmath}
\usepackage{multirow} 

\begin{document}

\title{The Origin of Persistent Shear Stress in Supercooled Liquids }

\author{Sneha Abraham and Peter Harrowell}
\affiliation{School of Chemistry,
  University of Sydney, Sydney N.S.W. 2006, Australia}

\begin{abstract}
{The persistence of shear stress fluctuations in viscous liquids is
 a direct consequence of the non-zero shear stress of the local potential minima which couples stress relaxation to transitions between inherent structures.
In simulations of 2D and 3D glass forming mixtures, we calculate the distribution of this inherent shear stress and demonstrate that the variance
 is independent of temperature and obeys a power law in density. The inherent
 stress is shown to involve only long wavelength fluctuations, evidence of the central role of the static boundary conditions in determining the residual stress left after the minimization of the potential energy. A temperature $T_{\eta}$ is defined to characterise
 the crossover to stress relaxation governed by binary collisions at high temperatures to low temperature relaxation dominated by the relaxation of the inherent stress. $T_{\eta}$ is found to coincide with the breakdown of the Stokes-Einstein scaling of diffusion and viscosity.}
\end{abstract}


\maketitle

\section{1. Introduction}
Many liquids, if cooled rapidly enough below their freezing point, will undergo a transition from fluidity to rigidity without any obvious
change in structure. The resulting glass states are ubiquitous and include many technologically important ceramics~\cite{ref1}, polymeric
materials~\cite{ref2} and metal alloys~\cite{ref3}.  The glass transition can be characterized by the rapid increase in the shear viscosity
 on cooling over a narrow temperature range, an increase that can be directly attributed to the enormous increase in the relaxation time of
 shear stress fluctuations.  While there is a large ongoing research effort focussed on understanding the generic slow down of particle dynamics
 in liquids on supercooling~\cite{ref4,ref5,ref6,ref7,ref8}, the explicit connection between the motions of atoms or molecules
 and the relaxation of stress in an amorphous material remains poorly understood.

As pointed out by Goldstein in 1969~\cite{ref14}, the slow relaxation of supercooled liquids can quite generally be attributed
 to the decreasing frequency, on cooling, of transitions between local minima in the potential energy surface over the N particle
 configuration space. The particle configuration associated with a local minima is referred to as an inherent structure.  We begin
 our argument by noting that for this slow process to be coupled to slow stress relaxation it is necessary that these inherent structures (IS) have
 non-zero shear stresses. Previously, Lacks~\cite{ref15} has described the nonlinear rheology of a liquid in terms of the stress in local minima
 due to the applied strain. Here we shall demonstrate that these minima are stressed even in the absence of external strain. The presence of these
 residual stress at equilibrium have featured in a number of recent studies.  In developing an algorithm to calculate stress relaxation based on transition rates about a representative network of inherent structures, Kushima et al
 ~\cite{yip} make explicit use of the fact that the inherent structures are stressed. Puosi and Leporini~\cite{leporini} have
 compared the relaxation of the stress in the parent liquid to the relaxation of the inherent stress. They found that after a crossover time $t^*$ the two autocorrelation functions converged, indicating that
 stress relaxation beyond $t^*$ was dominated by the relaxation of inherent structure stress. They went on to demonstrate that the plateau modulus $G_p$,
 related to the variance of the inherent stress by $G_{p} \approx V<\sigma_{xy}^{2}(IS)>/T$, provided a general energy scale for the slow alpha
 relaxation process. Bailey et al~\cite{bailey} have examined the statistics of the magnitude of the change in shear stress (along with that of energy and pressure) as a supercooled liquid moves between inherent structures. The authors of each of these studies~\cite{yip,leporini,bailey} appeared to consider the existence of the inherent structure (IS) stress as self evident with little
 discussion required. While acknowledging the logic that very slow stress relaxation implies the existence of IS stress, we believe that the factors
 that determine the magnitude of this residual stress, let alone the physical explanation of how the local minima come to be stressed in the first place, are far from obvious. Given the fundamental role played by the inherent structure stress in the large increase in shear viscosity on cooling and the apparent importance of the variance of this stress $<\sigma_{xy}^{2}(IS)>$ in establishing the kinetic energy scale, there is a clear need to characterise and understand the stresses inherent in the liquid energy landscape.

In this paper we begin by characterising the inherent shear stress in simulations of glass forming mixtures in 2D and 3D and, then, demonstrate that the long time tail of the shear stress autocorrelation arises through the slow relaxation of this inherent stress. We establish that the variance of the IS stress is independent of temperature but strongly dependent on density. Specifically, we show that, as
the density is decreased, the local potential minima become unable to sustain any persistent stress~\cite{ref9} (and, hence, support
a glass transition), in a manner that can be explicitly connected to the interactions between atoms. In Section 5, we consider the origin of the stressed states and present evidence that they are the result of boundary constraints on the rigid inherent structures. We argue, in Section 6, that the essential mechanical transition experienced by a liquid on cooling occurs
at a temperature well above the glass transition temperature and corresponds to the crossover from the high temperature liquid
to the viscous liquid, the latter characterised by stress relaxation dominated by the residual stress.

\section{2. Models, Algorithms and the Ensemble Dependence of the Green-Kubo Expression}

We work with constant NVT molecular dynamics simulations. The equations of motion are integrated using the velocity-Verlet algorithm~\cite{ref34} and
the temperature is constrained using the Nose-Hoover thermostat~\cite{ref34}. We have chosen two well studied models of glass forming binary alloys, one
in 2D~\cite{donna} and one in 3D~\cite{ref17}. In 2D we have used the binary equimolar soft disk mixture with an interaction potential between species
i and j given by $\phi_{ij}(r)=\epsilon (\frac{a_{ij}}{r})^{12}$  where $a_{11}=1.0$, $a_{12}=1.2$ and $a_{22}=1.4$.  The mass of the two
 species are equal. In 3D we have used another binary equimolar mixture, this time interacting via a Lennard-Jones interaction
$\phi_{ij}(r)/4\epsilon= (\frac{a_{ij}}{r})^{12}-(\frac{a_{ij}}{r})^{6}$ where $a_{11} = 1.0$, $a_{12}=1.1$ and $a_{22}=1.2$.
The mass of particle 2 is twice that of particle 1. These potentials have been truncated at a distance $r_{c} = 4.5a_{22}$ and
the potential shifted so that $\tilde{\phi}_{ij}(r)=\phi_{ij}(r)-\phi_{ij}(r_c)$. The component $\sigma_{ab}$ of the stress tensor
(where $a$ and $b$ each correspond to one of the Cartesian axes $(x,y,z)$) is given by

\begin{equation}
\label{stress}
\sigma_{a b}=\sum_{i=1}^{N}m_{i}u_{i a}u_{i b} + \frac{1}{2}\sum_{i=1}^{N}\sum_{j\neq i =1}^{N} r_{ij}^{a}r_{ij}^{b}F_{ij}
\end{equation}

\noindent where $u_{ia}$ and $r^{a}$ refer to the component of the particle $i$'s velocity and the projection of the vector $\vec{r}$, respectively, along the Cartesian axes $a$. The force $F_{ij}$ is given by

\begin{equation}
F_{ij}= \frac{-1}{r_{ij}} \frac{d\phi_{ij}}{dr_{ij}}
\end{equation}

\noindent The shear stress is the off diagonal component $\sigma_{xy}$ while the pressure $P= (\sigma_{xx}+\sigma_{yy}+\sigma_{zz})/3$. When calculating the inherent structure shear stress or pressure, Eq.~\ref{stress} is used but with all velocities set to zero.

The following reduced units are used
throughout the paper: length, $L = L/a_{11}$; time, $t=t/(m_{1}a_{11}^{2}/\epsilon)^{1/2}$  ; temperature $T = k_{B} T/\epsilon$,
pressure, $P=Pa_{11}^{3}/\epsilon$  (also stress) and energy, $E = E/\epsilon$. Unless otherwise indicated, the following
number density $\rho$ and number of particles N were used: $\rho=0.7468$ and N=1024 (2D) and $\rho=0.75$ and N=1024 (3D). At these densities, the freezing temperatures for the respective crystal phases are $T_{f}(3D)=0.6$ \cite{tox} and $T_{f}(2D)\approx 0.70$ \cite{donna}.  A trajectory of configurations corresponding to the local potential energy minima is generated by carrying out a
conjugate gradient minimization of the potential energy from configurations generated at time intervals of $dt = 0.003\tau_{\alpha}$ along
a given trajectory. ($\tau_{\alpha}$ is the stress relaxation time obtained from a stretched exponential fit to the long time
tail of the shear stress autocorrelation function.

In the linear response regime, the shear viscosity $\eta$ can be written as a time integral over the equilibrium shear stress
autocorrelation function~\cite{ref12},

\begin{equation}
\label{GK}
\eta=\frac{V}{k_{B}T}\int_{0}^{\infty}dt<\sigma_{xy}(0)\sigma_{xy}(t)>
\end{equation}

\noindent Note that
this stress refers to the instantaneous shear stress of the entire system.
We can rewrite Eq.~\ref{GK} as $\eta=G_{\infty}\tau$  where the infinite frequency shear modulus, $G_{\infty}$, given by

\begin{equation}
\label{ginfty}
G_{\infty} = \frac{V}{k_{B}T}<\sigma_{xy}^{2}>,
\end{equation}

\noindent and $\tau$ is the relaxation time of the shear stress fluctuations. $G_{\infty}$ exhibits only modest temperature
dependence~\cite{ref13}, so that the enormous increase in shear viscosity originates in an equally enormous growth of $\tau$.

The validity of Eq.~\ref{GK} depends on the choice of ensemble over which the average, indicated by the angle brackets, is taken.
It is, perhaps, trivial to observe that choice of an ensemble in which the shear stress was constrained to be constant would,
by suppressing stress fluctuations, render Eq.~\ref{GK} invalid. What is less obvious is that the
derivation of Eq.~\ref{GK} implies that the equilibrium stress fluctuations are subjected to a constraint corresponding
to a fixed cell shape. The argument goes as follows. Consider a nonequilibrium calculation to determine the shear viscosity in which a
sample is subjected to a (small) and rapidly applied fixed shear strain. The relaxation time of the resulting shear stress is, in
linear response, simply proportional to the shear viscosity. It is clear that in such a set up the applied shear strain
must be held fixed since, if it was not, the stress would be provided an alternate relaxation path
(i.e. the elastic restoration of the zero strain state) that has nothing to do with the viscosity of the liquid.
We arrive at Eq.~\ref{GK} by application of Onsager's regression hypothesis~\cite{onsager} in which the relaxation
of the stress perturbed by the applied strain is replaced by the relaxation of the stress arising from equilibrium fluctuations.
The requirement that the applied strain (i.e the box shape) remains fixed, however, remains when considering the relaxation of
the equilibrium fluctuations for the same reason as applied in the nonequilibrium case. To summarise, the use of Eq.~\ref{GK}
to calculate the shear viscosity requires that the simulations are carried out in a cell of fixed shape.
We are not suggesting that the shear viscosity itself depends on the choice of ensemble, but rather just the fluctuation-dissipation relation
used to evaluate it. As the reader may have already guessed from this rather extended discussion, we shall return to
the role of boundary conditions and stress relaxation later in this paper.

\section{3. The Contribution of the Inherent Structure Stress to Stress Relaxation}

\begin{figure}[!htb]
\centering
\includegraphics[scale=.5]{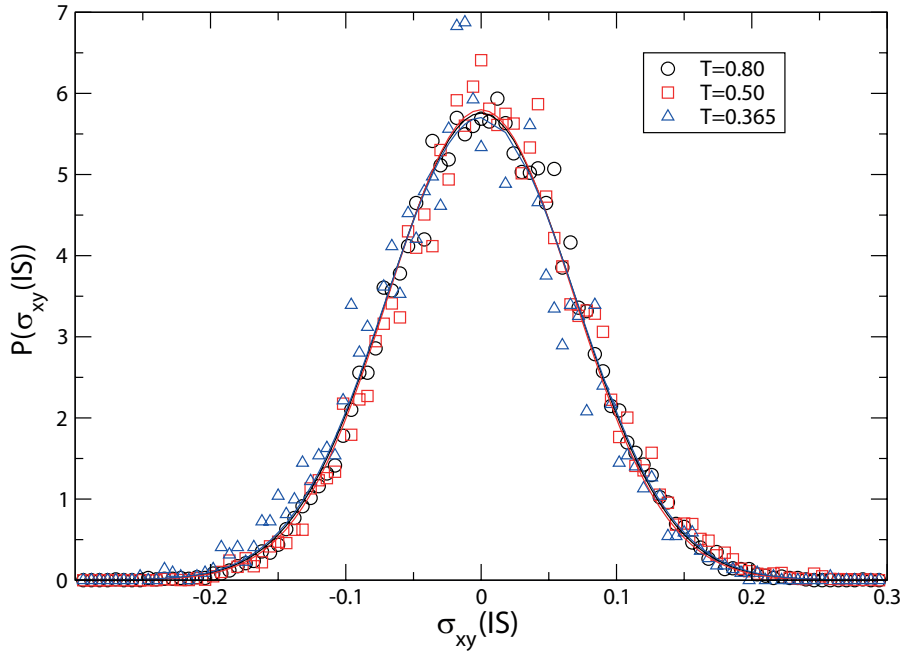}
\caption{The distribution of the inherent structure shear stress in the 2D soft disc liquid mixture for a range of temperatures. The three curves are Gaussian fits to the data for the three different temperatures.}
\label{figa}
\end{figure}

\begin{figure}[!htb]
\centering
\includegraphics[scale=.5]{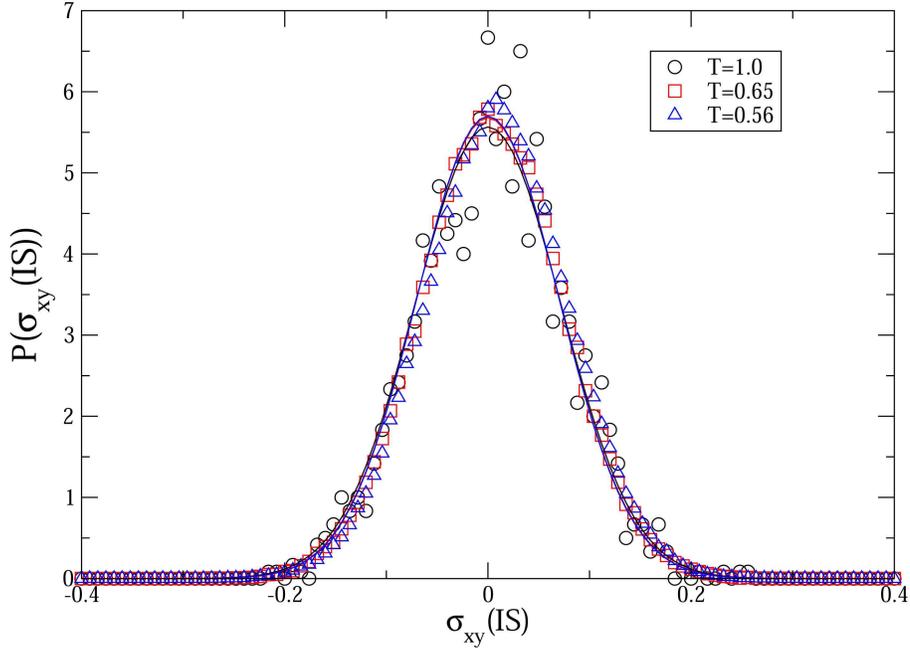}
\caption{\label{figb}  The distribution of the inherent structure shear stress in the 3D-LJ liquid for a range of temperatures. The three curves are Gaussian fits to the data for the three different temperatures.}
\end{figure}

\begin{figure}[!htb]
\centering
\includegraphics[scale=.5]{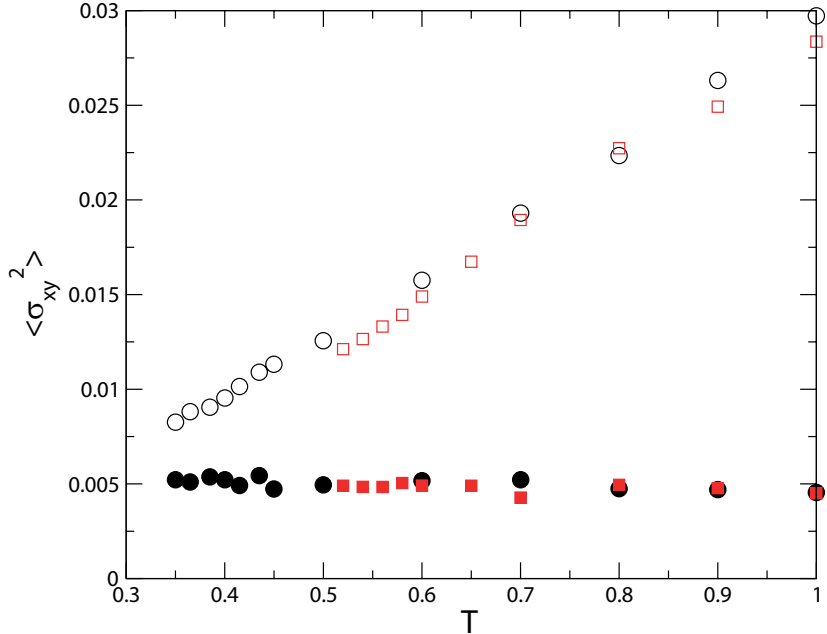}
\caption{\label{fig6}  The variance of the shear stress $<\sigma_{xy}^{2}>$ vs T for
the parent liquid (open symbols) and the inherent structures (filled symbols) for the 2D (circles) and 3D (squares) liquids.}
\end{figure}

\begin{figure}[!htb]
\centering
\includegraphics[scale=.7]{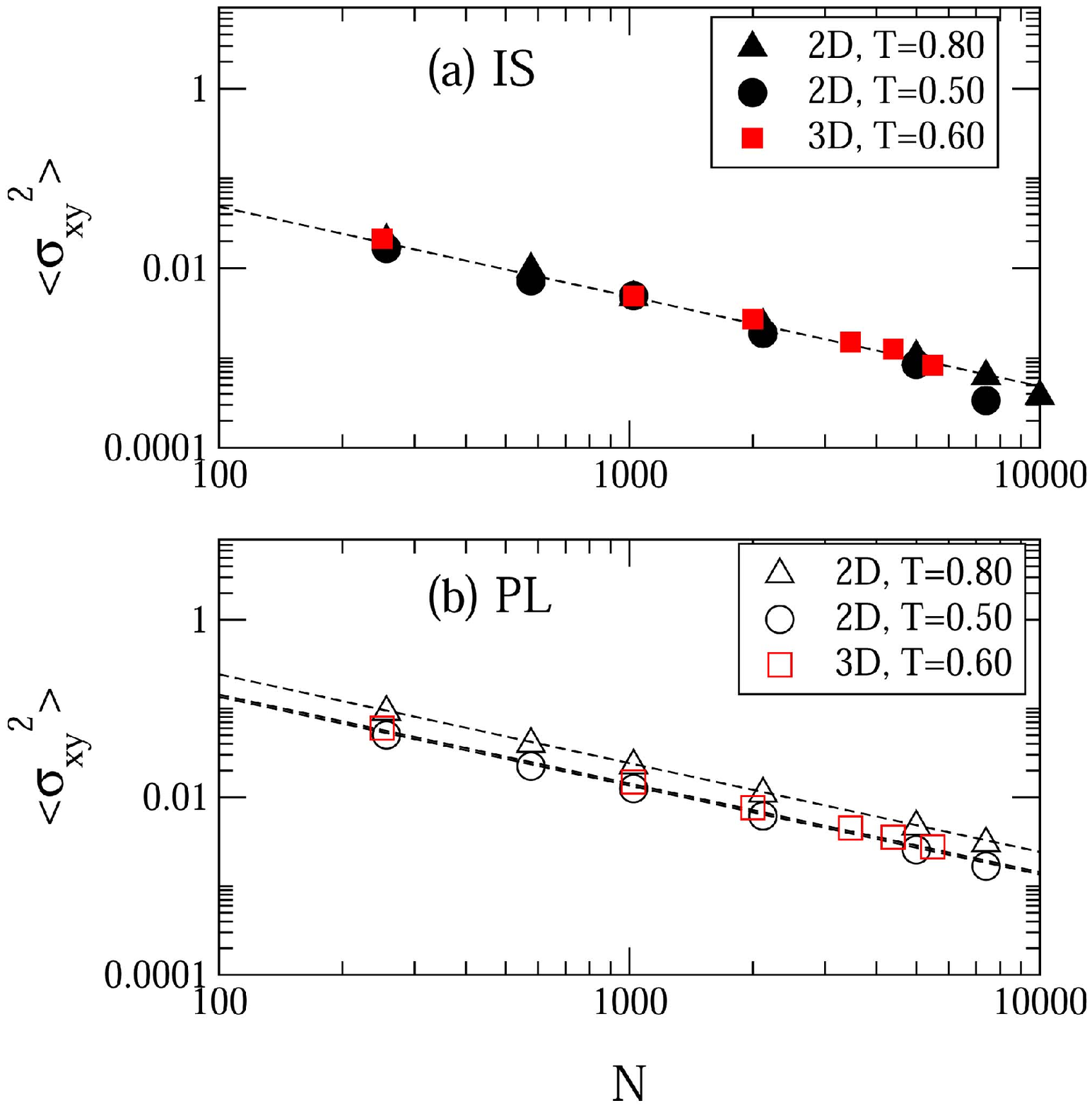}
\caption{\label{figc}  A log-log plot of the shear stress variance vs the system size $N$ for a) the inherent structure contribution and b) the parent liquid. The dashed lines all have a slope of one. Data is presented from the 2D and 3D models, as indicated.}
\end{figure}

   The distributions of the shear stress $\sigma_{xy}$ in inherent structures of the 2D and 3D liquids are
plotted in Figs.~\ref{figa} and~\ref{figb}, respectively. The fact that the inherent structures can sustain non-zero shear stresses identifies these minima as solids. We find that the distributions are well
described by a single Gaussian. A striking feature of both sets of IS stress distributions is their independence of temperature.
While the variance of the shear stress $<\sigma_{xy}^{2}>$ in the parent liquid decreases linearly with temperature
in both the 2D and 3D-LJ mixtures (as shown in Fig. \ref{fig6}), the analogous variance in the inherent structures
is essentially independent of temperature. The variances of the shear stress in both the parent liquid and the inherent structures varies with the
number of particles N as $<\sigma_{xy}^{2}> \propto 1/N$ as shown in Fig.~\ref{figc}. The absence of any temperature dependence of the stress distribution is surprising,
given that there is a strong dependence of the inherent structure energy on temperature. Given its temperature independence, the inherent stress distribution can be regarded as a density of states - an alternate one in which local minima are identified by their shear stress instead of their energy.

If extrapolated on to lower temperatures, the behaviour depicted in Fig.~\ref{fig6} must see the two variances intersect
 at a temperature between 0.25 and 0.3 for either liquid. Since it is impossible for the inherent contribution to
 exceed that of the parent liquid (because $<\sigma^{2}_{xy}(PL)> \approx <\sigma^{2}_{xy}(IS)> + <\sigma^{2}_{xy}(vib)>$, the latter contribution arising from transverse phonons), one of two things must happen. Either i)  the IS variance develops a temperature
dependence and vanishes with T or ii) the parent liquid stress loses its T dependence and we are left with a non-zero
 residual stress at T = 0.  Case i), we suggest, represents the equilibrium behaviour of the liquid while case ii) corresponds
 more closely to what would actually be seen as the liquid drops out of equilibrium and is kinetically arrested. In case i), the
 Green-Kubo expression for the $G_{\infty}$ would continue to apply all the way to T = 0. In case ii), however, Eq.~\ref{GK} no
 longer provides a correct description of the relationship between stress fluctuations and the elastic modulus since it assumes
that the average shear stress is zero. As has been discussed by Ilyin et al~\cite{ref18} and Williams~\cite{ref19}, the
correct statistical expression for $G_{\infty}$ in case ii) is that derived by Squire et al~\cite{ref20} for a rigid solid.

\begin{figure}[!htb]
\centering
\includegraphics[scale=.7]{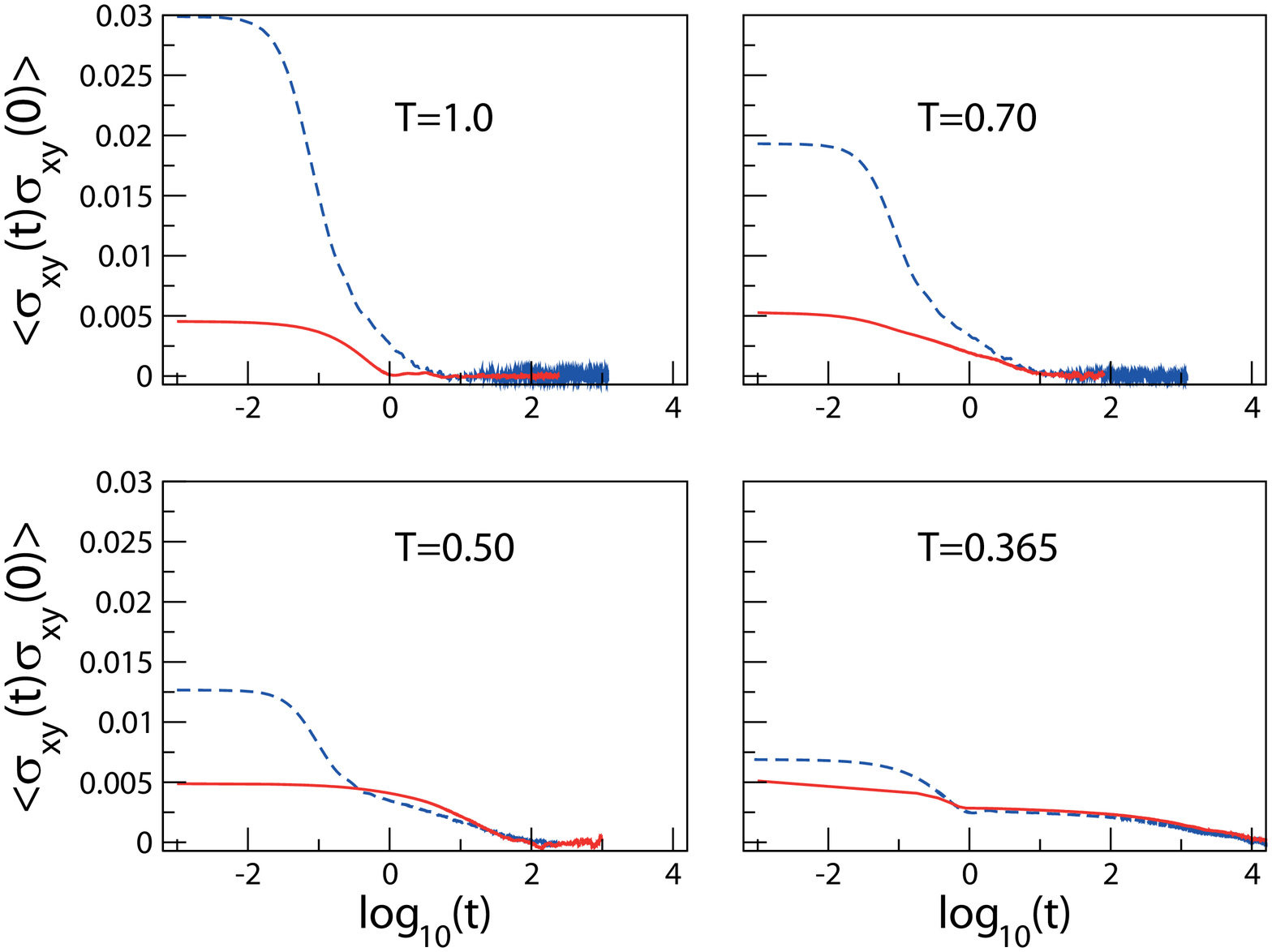}
\caption{\label{fig1}  The shear stress autocorrelation function $<\sigma_{xy}(0)\sigma_{xy}(t)>$ for the
parent liquid (dashed line) and IS stress (solid line) for the 2D liquid at T=1.0, 0.70, 0.50 and 0.365.}
\end{figure}

\begin{figure}[!htb]
\centering
\includegraphics[scale=.7]{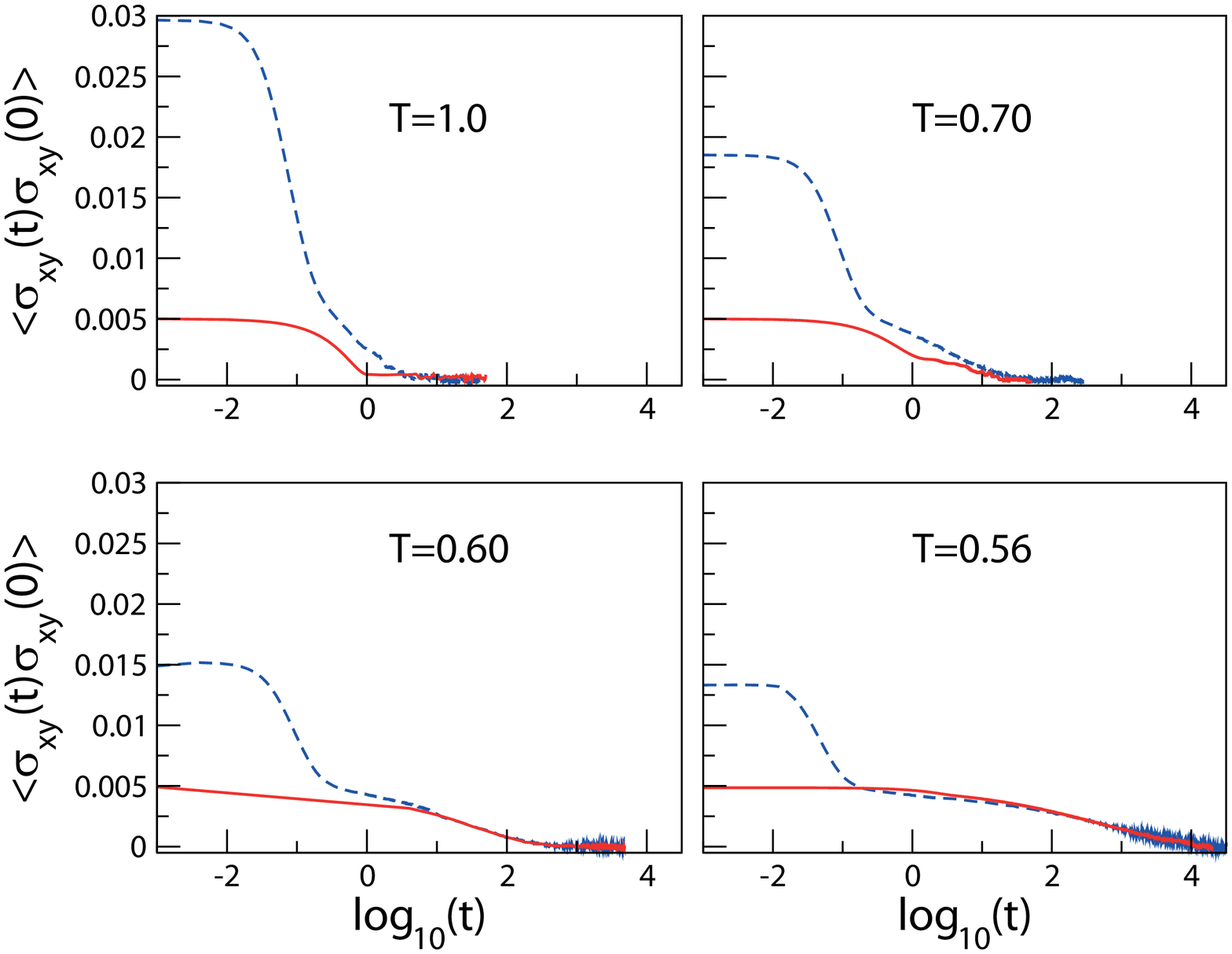}
\caption{\label{fig2} The shear stress autocorrelation function $<\sigma_{xy}(0)\sigma_{xy}(t)>$ for
the parent liquid (dashed line) and IS stress (solid line) for the 3D liquid at T=1.0, 0.70, 0.60 and 0.56.}
\end{figure}

Having quantified the magnitude of the residual shear stress associated with the inherent structures, we
now consider its relaxation dynamics.  In Figs.~\ref{fig1} and~\ref{fig2} we plot the
shear stress autocorrelation functions for our 2D and 3D liquids, respectively. For both systems,
we can see that the long time tail of the stress autocorrelation that develops at low temperatures is
completely accounted for by the relaxation of the inherent structure stress. With decreasing temperature,
the IS contribution to the stress autocorrelation function makes an increasingly larger contribution to
the total stress fluctuation and, crucially, the life time of the IS stress fluctuations grows rapidly. In contrast, we find that the relaxation time of the fast component of stress (the difference between the two curves in each panel of Figs. ~\ref{fig1} and ~\ref{fig2}) show negligible temperature dependence. As can be seen in Figs. ~\ref{fig1} and ~\ref{fig2}, the height of the plateau in the shear stress
autocorrelation function at large supercoolings is equal to $ <\sigma^{2}_{xy}(IS)> $.

\begin{figure}[!htb]
\centering
\includegraphics[scale=.5]{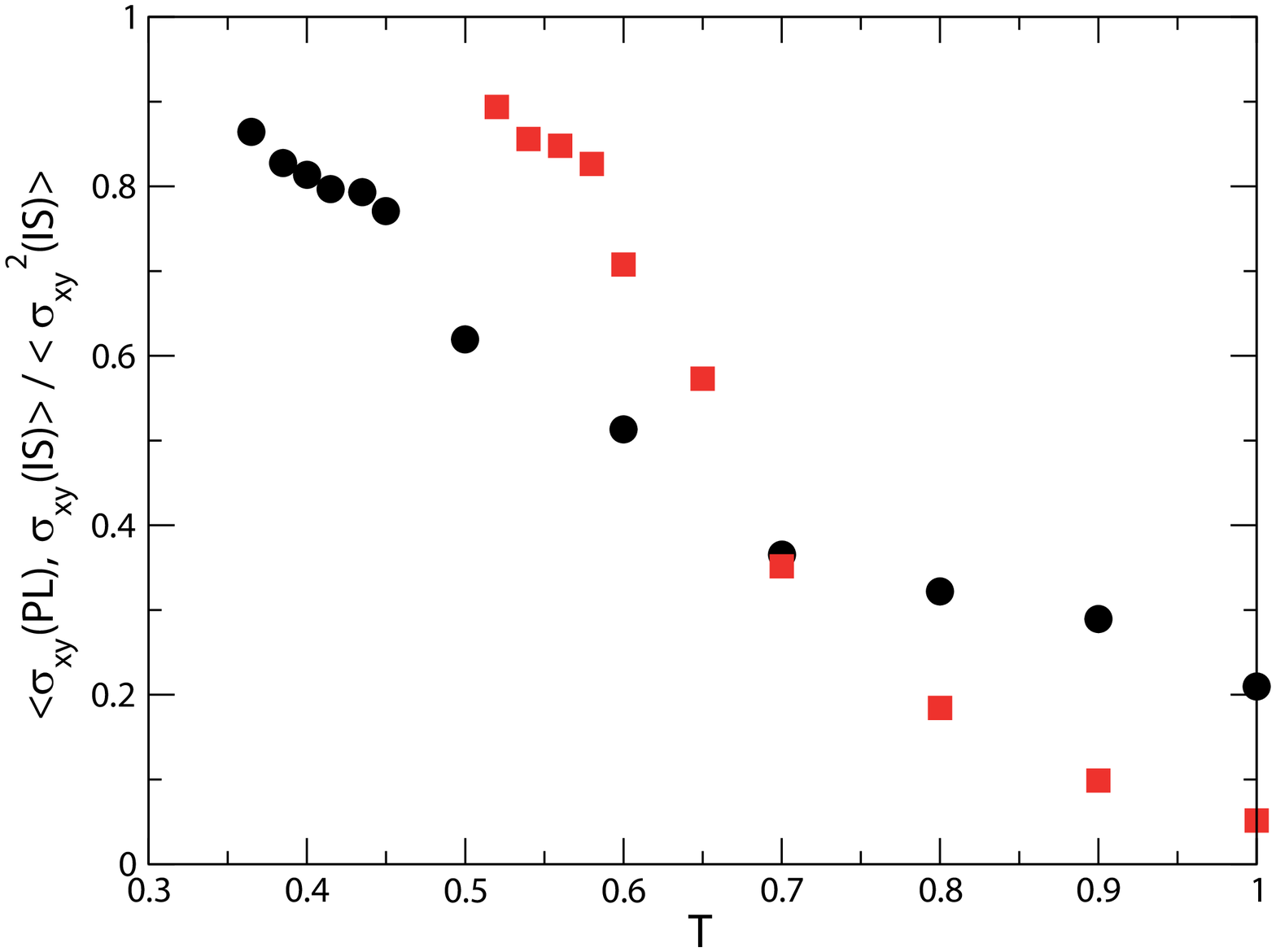}
  \caption{\label{fig7}  The relative covariance $<\sigma_{xy}(PL)\sigma_{xy}(IS)>/ <\sigma_{xy}^{2}(IS) >$ vs T
for the 2D (circle) and 3D (square) liquids. {\it PL} refers to the parent liquid.}
\end{figure}

In order to establish whether there was any direct correlation between the stress fluctuations in the parent
liquid and those observed in the inherent structures obtained after carrying out energy minimizations,
we examined the covariance between the shear stress. This covariance is between the instantaneous shear stress in
the parent liquid and the shear stress in the inherent structure generated from that specific parent liquid configuration.
This covariance (normalised by the variance from the inherent structures) is plotted in Fig.~\ref{fig7} for the 2D and 3D-LJ liquids.
In both cases we see little correlation at high T but, as the temperature is lowered, we observe a clear increase in
correlation, tending towards a value of one.  This is additional evidence that the parent liquid fluctuations actually
do reflect the same residual stress as that obtained by the minimization of the potential energy.

\section{4. Density Dependence of the Inherent Structure Stress}

\begin{figure}[!htb]
\centering
\includegraphics[scale=.7]{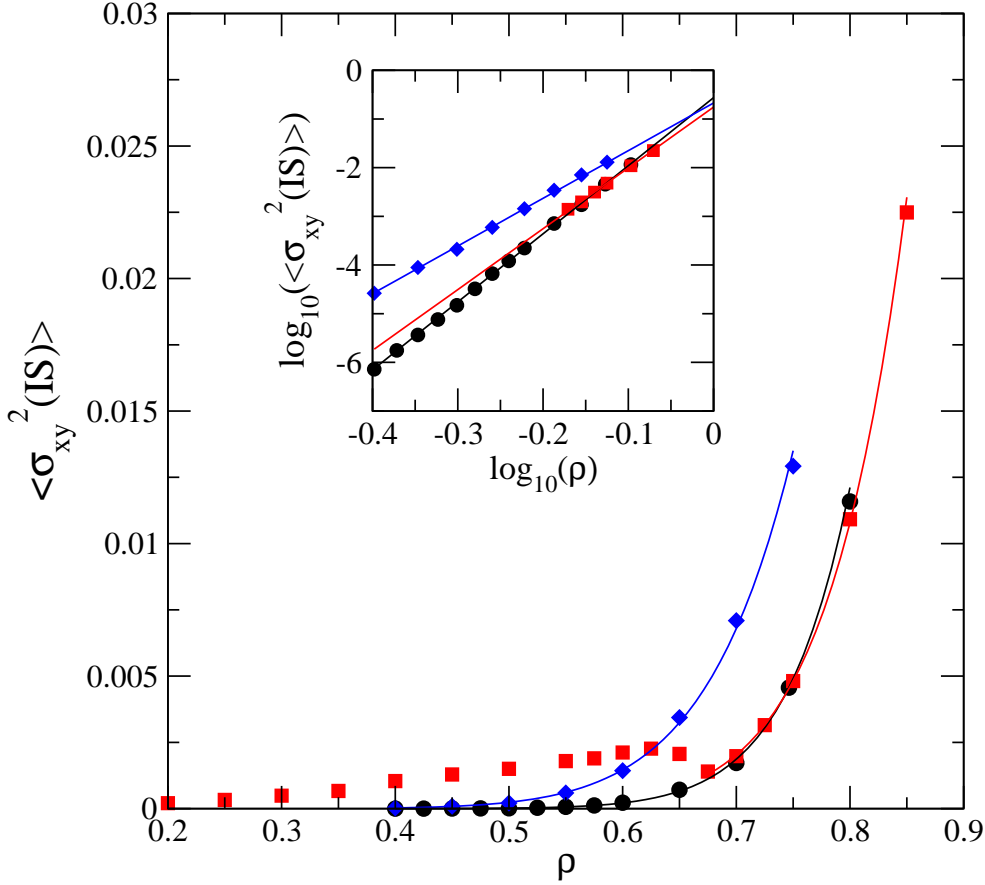}
\caption{\label{fig3}  The variance of the IS shear stress $<\sigma_{xy}^{2}(IS) >$ vs density for the 2D (circles),
3D (squares) and the 3D liquid with a purely repulsive potential (diamonds). Inset: Log-log plot of the same
data except that, for the 3D-LJ, only data for $\rho > 0.7$ are included.}
\end{figure}

Since an inherent structure will only be able to maintain a shear stress fluctuation whose magnitude
is less than the yield stress of that inherent structure, we might expect that the residual stress will
vanish with decreasing density in a manner analogous to the (un)jamming transition of granular material~\cite{ref9}.
In Fig.~\ref{fig3} we plot the behaviour of the IS stress variance $<\sigma_{xy}^{2}>$  as a function of
density for both our 2D and 3D-LJ models and the 3D model with only the repulsive $r^{-12}$ term retained. While the
expected strong density dependence of the residual shear stress is observed, we do not observe any sign of the singular (un)jamming
transition reported for particles with truncated potentials~\cite{ref9}. For the inverse power repulsions in 2D and 3D, we
find that $<\sigma_{xy}^{2}> \sim \rho^{\gamma}$ over the entire range of density $\rho$ with values of $\gamma$ of 13.98 and 9.78, respectively.
For an inverse power law potential of the form $r^{-n}$, a simple scaling argument~\cite{ref21} predicts that IS stress variance
will vary with the density $\rho$ as  $<\sigma_{xy}^{2}> \sim \rho^{\gamma}$, where

\begin{equation}
\label{gamma}
\gamma = 2(1+n/d)
\end{equation}									

\noindent and d is the spatial dimension. The predicted values of $\gamma$ for the $r^{-12}$ repulsions in 2D and 3D, respectively,
are 14 and 10, in good agreement with the observed exponents. Klix et al~\cite{ref22} have recently reported measurements
of the shear modulus of an amorphous colloidal suspension in 2D as a function of the coupling parameter that bears
an interesting correspondence with the curves in Fig.~\ref{fig3}.

\begin{figure}[!htb]
\centering
\includegraphics[scale=.7]{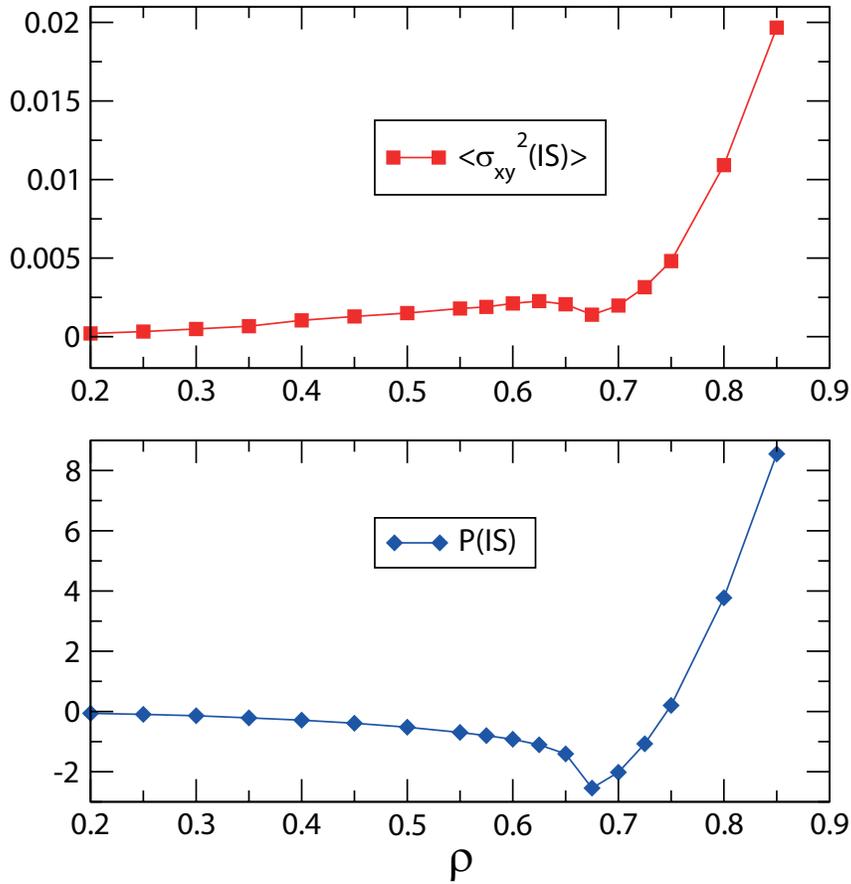}
\caption{\label{talk}  Upper frame. The variance of the inherent structure shear stress as a function of density
for the 3D-LJ liquid. Lower frame. The potential contribution to the pressure from the inherent structures of the 3D-LJ liquid. }
\end{figure}

\begin{figure}[!htb]
\centering
\includegraphics[scale=.7]{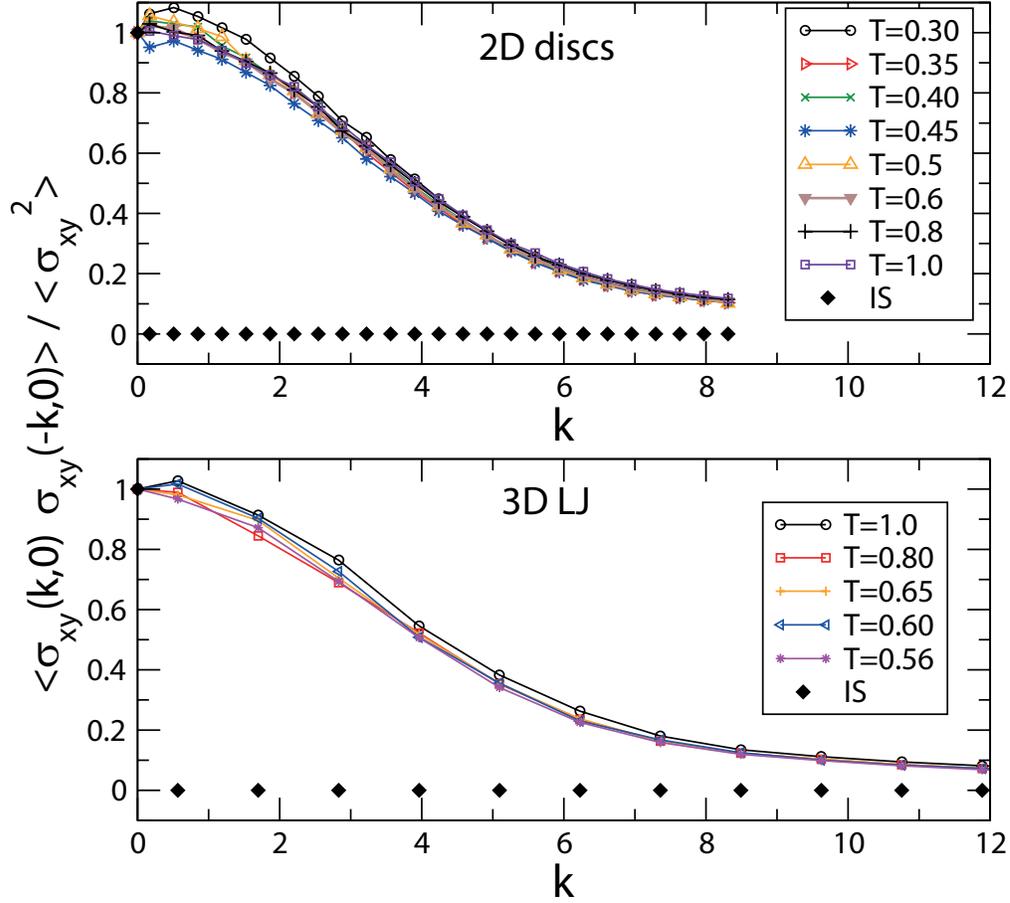}
\caption{\label{fig4}  The dependence of the variance of Fourier components of the shear stress on the
magnitude k of the wavevector for the 2D (top) and 3D-LJ (bottom) liquids. The data from the parent liquid
is presented for a range of temperatures and the curve represents a Gaussian fit. The k-dependent variance of
the IS stress is shown as black diamonds in both plots with nonzero values only at k = 0.}
\end{figure}

For the 3D-LJ mixture, the average pressure of the inherent structures passes through zero at a reduced density $\sim 0.75$
and exhibits a minima at $\sim 0.675$, the same density at which the variance of the IS stress exhibits a local minimum
(see Fig.~\ref{talk}). At densities above $\rho = 0.675$, the IS can be characterized as repulsive glasses with compressive
stress, while below this threshold we have attractive glasses under tension~\cite{ref23}. Below the threshold density, the inherent structures show evidence of cavitation and, as the density decreases, cease to be of relevance with respect to the properties of the parent liquid. This crossover results in the
non-monotonic density dependence of $<\sigma_{xy}^{2}>$  seen in Fig.~\ref{fig3}. For densities greater than 0.675, the
LJ mixtures exhibits a surprising similarity to the 2D disc mixture. On fitting $<\sigma_{xy}^{2}>$  to $\rho^{\gamma}$ we
find an exponent of $\gamma \sim 12.49$ which, if inserted into Eq.~\ref{gamma} would give us an effective inverse power
law repulsion of $r^{-15.73}$.  If, following Pedersen et al~\cite{ref24}, we determine the effective inverse power law
exponent n through the relationship {\it{W = (n/3)U}}  between the viral W and the potential energy U of the 3D-LJ liquid,
we find that our LJ mixture can be modeled using an inverse power law interaction of the form $r^{-15.27}$, in reasonable
agreement with the effective interaction obtained from the density dependence of $<\sigma_{xy}^{2}>$ . The near coincidence
of the LJ and soft disk curves in Fig.~\ref{fig4} appears to be a bit of serendipity, arising as a consequence of Eq.~\ref{gamma}
and the fact that increasing n or decreasing d has a similar effect on the exponent $\gamma$.

\section{5. Why are the Inherent Structures Stressed?}

Given the essential role played by the IS stress in coupling stress relaxation to the frequency of transitions between the inherent
 structures at equilibrium, we need to understand why the local potential minima are stressed in the first place. As we shall show, the length scale associated with the inherent structure stress is central to answering this question. To this end , we consider the variance of the
various wavevector components of the shear stress fluctuations
$\langle \vec{\sigma}_{\alpha \beta}^{k}(0)\vec{\sigma}_{\alpha \beta}^{-k}(0) \rangle $ ,  where (following ref. ~\cite{ref25})

\begin{equation}
\label{kdep}
\vec{\sigma}_{\alpha \beta}^{k}(t)=\sum_{i=1}^{N}m_{i}u_{i \alpha}u_{i \beta}e^{-i\vec{k}\cdot \vec{r}_{i}(t)} + \frac{1}{2}\sum_{i=1}^{N}\sum_{j\neq i =1}^{N} \frac{r_{ij}^{\alpha}r_{ij}^{\beta}}{r_{ij}^{2}} \Phi_{\vec{k}}(\vec{r}_{ij})e^{-i\vec{k} \cdot \vec{r}_{i}(t)}
\end{equation}
 			
\noindent with $\Phi_{\vec{k}}(\vec{r}_{ij}) \equiv \frac{r_{ij}\phi'(r_{ij})}{i\vec{k} \cdot \vec{r_{ij}}}[ e^{i\vec{k} \cdot \vec{r}_{ij}(t)}-1]$.
Here $\alpha,\beta =x,y,z$ and $\phi ' = \nabla \phi$. As shown in Fig.~\ref{fig4},  while the parent liquid exhibits a
continuous variation over the wavevector k, the IS stress variance resembles a delta function centered at k = 0. The energy minimization has zeroed the shear stress at all wavevectors save $k=0$, i.e. those fluctuations that span the simulation cell. Since the relaxation of even the $k=0$ mode could be achieved by the energy minimization {\it{if}} the shape of the box was allowed to change, we conclude that the inherent stress reflects the constraint imposed on the energy minimization by the fixed shape of the simulation cell. At this point, we remind the reader  of our discussion in Section 2 regarding the requirement that the simulation cell shape be held fixed in order that we can associate the shear viscosity (i.e. Eq.~\ref{GK}) with the relaxation of the stress fluctuations. The influence of the fixed cell shape, in other words, cannot be simply dismissed as an artefact arising from an arbitrary choice of boundary condition.

While the inherent stress is, by construction, that which remains after energy minimization, it is useful, we suggest, to equate this residual stress with that stress incurred by generating an extended (aperiodic) solid (i.e. an inherent structure) in a cell of fixed shape. Having an intrinsic shape is, after all, a defining feature that differentiates a solid from a fluid. The existence of the inherent structure shear stress, then, can be regarded
as the signature of the solid-like nature of the local minima.

The onset of rigidity in supercooled liquids has been described by a number of workers as arising from the formation of local rigid aggregates whose size increases on cooling~\cite{ref28,ref29}. This picture provides an appealing account of the threshold wavelength of shear waves in a liquid, above which the modes become overdamped~\cite{nelson}. The results presented in this paper, however, suggest an alternative picture, one in which the viscosity increase is due to the increasing lifetime of the extended system-spanning rigid solids that constitute the local potential minima. This alternate picture is consistent with the recent report of long range stress correlation in a simulated glass forming liquid~\cite{ref27}. While our results underscore the importance of the long correlations associated with stressed solids, we note that reorganization events responsible for transition between inherent structures are localised and so do introduce a length scale into the stress relaxation. Whether the cluster picture can be made consistent with the role of the stress IS described here will require detailed study of the spatial character of the changes in IS stress due to transitions between the potential minima.

\section{6. Crossover in the Mechanical Response of a Liquid}

The result presented in this paper leave us with the following picture of the transition from fluidity to rigidity. The solid state is always present in the liquid, in the sense that the liquid, at any temperature, explores a configuration space that strongly overlaps that of the aperiodic solids we call inherent structures. (This is, of course, very different from the case of crystalline minima whose density in the configuration space is so much smaller than that of the amorphous inherent structures that they typically make no contribution to the liquid state.) The mechanical correlation length of the local minima is effectively infinite in the sense that a non-zero shear stress can be generated through the sample by the imposition of a static boundary strain. It is on this point that we differ from the commonly presented picture of the viscous liquids in which solid-like behaviour is restricted to short length scales~\cite{ref29}. The viscosity decreases with increasing temperature, we suggest, not through the decrease of some stress correlation length scale, but due to the decreasing lifetime of the individual inherent structures and the growing contribution from the short lived stress fluctuations of shear vibrational modes.

\begin{figure}[!htb]
\centering
\includegraphics[scale=.7]{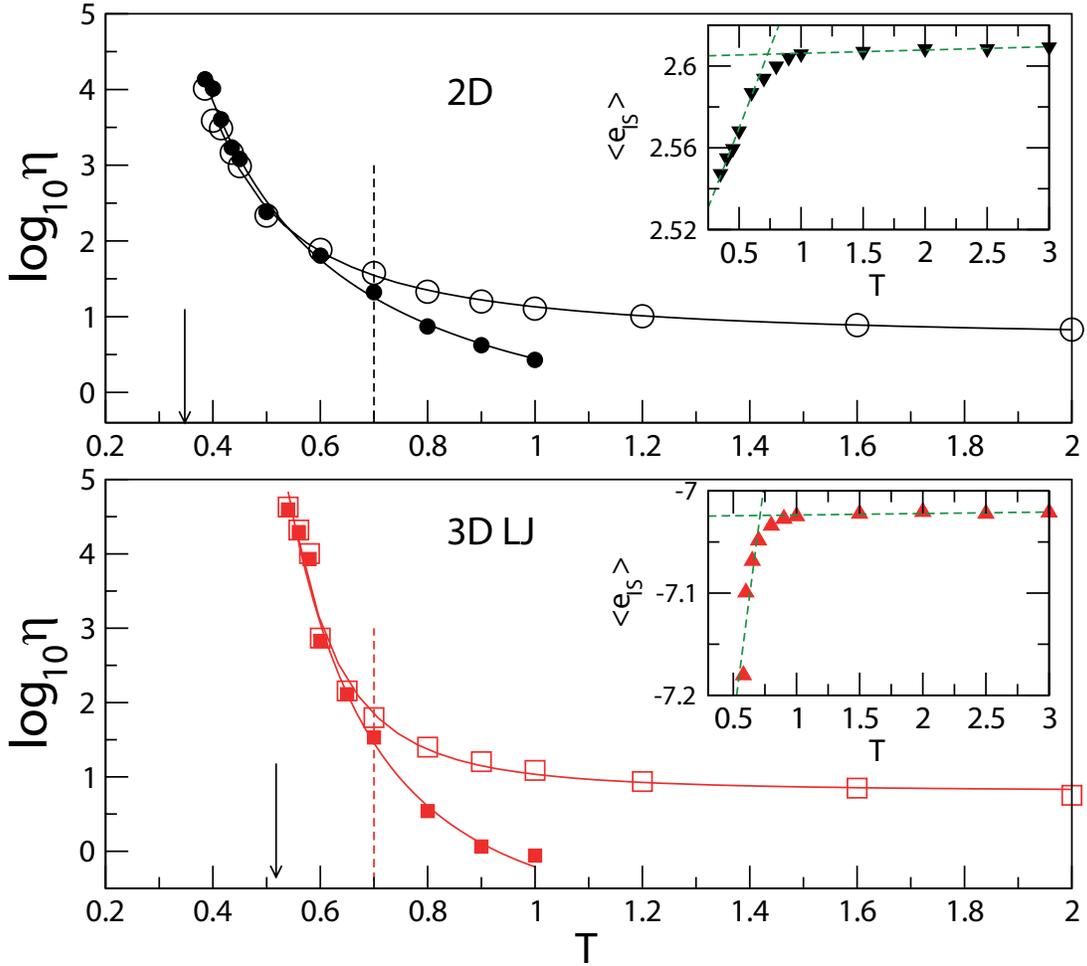}
\caption{\label{fig5}  Log of viscosity, $log_{10}$ vs T showing the parent liquid viscosity $\eta$ (open symbols) and
the IS contribution $\eta_{IS}$ (filled symbols) for the 2D (circle) and 3D LJ (square) liquids.
The curves are fits to the function proposed by Mauro et al~\cite{ref31}. The crossover temperatures,
corresponding to the point where $\eta_{IS}/\eta =0.5$, are indicated by the dashed lines.
The arrows indicate $T_{c}$ (as defined in the text). Inserts:
The average inherent structure energy per particle $<e_{IS}>$ vs the temperature of the parent liquid.
The landscape temperature $T_o$ is estimated as the crossing point of the dashed lines.}
\end{figure}

The crossover in stress relaxation from that dominated by uncorrelated collisions at high temperatures to that
dominated by the residual stress in the inherent structures represents the fundamental passage of a liquid from
fluid to rigid behaviour on cooling. We can locate this crossover with a temperature $T_{\eta}$, defined as the
temperature at which the IS contribution to the viscosity corresponds to half the total viscosity (see Fig.~\ref{fig5}).
For comparison, the glass transition temperatures $T_{g}$ are estimated to be 0.26 and 0.46 for the 2D and 3D liquids,
respectively, based on an empirical fitting function~\cite{ref31}.  As indicated in Fig.~\ref{fig5}, $T_{\eta} \sim 0.7$
for both the 2D and 3D-LJ liquids (at the densities selected), values considerably higher than both $T_{g}$ and the mode
coupling temperature $T_{c}$ (obtained from fitting the diffusion constant $D \sim (T-T_{c})^{\alpha}$~\cite{ref32} and
indicated with arrows). $T_{\eta}$ is similar, in magnitude, to the landscape crossover temperature associated with the T
dependence of the average landscape energy~\cite{ref33} (see Fig.~\ref{fig5} insert). Based on a previous calculation of the
soft disk mixture phase diagram~\cite{donna}, we estimate that $T_{\eta}$ lies just below the freezing point $T_f$ for the
disk mixture at $T_{f}= 0.7468$. There is no reason why $T_{\eta}$ must be less than $T_{f}$. Whether there is a general
relation between $T_{\eta}$ and $T_{f}$ is an interesting open question.

\begin{figure}[!htb]
\centering
\includegraphics[scale=.7]{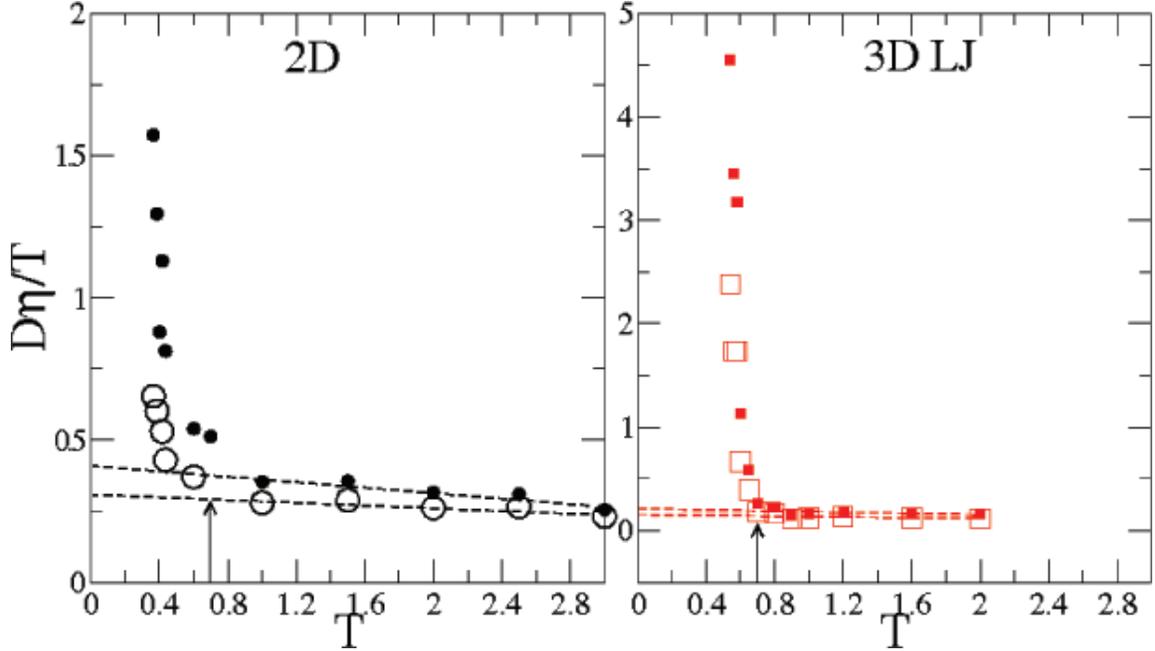}
 \caption{\label{SE}  The quantity $D\eta/T$ vs temperature for the 2D (left frame) and 3D-LJ (right frame) liquids with the crossover temperature $T_{\eta}$ indicated by an arrow in each plot.For each liquid the filled and open symbols correspond to the use of the small or large particle diffusion constant, respectively. The dashed lines correspond straight line fits to the high temperature data. }
\end{figure}

The mechanical crossover temperature $T_{\eta}$ clearly does not describe the transition from a liquid to a glass. Instead, it
marks the crossover from the high temperature liquid to the viscous liquid in simple liquids. In this sense, $T_{\eta}$ shares much in common with the crossover recently proposed by Brazhkin et al~\cite{brazhkin} based on the temperature at which the stress relaxation time equals the minimum period of the transverse waves. Without the input from the underlying solid minima, the viscosities of liquids would be limited to values
no greater than $G_{\infty}\times\tau_{fast}$ where $\tau_{fast}$ is the relaxation time of the fast component of stress fluctuation
associated with the damped transverse phonons in the liquid.  The explicit mechanical definition of $T_{\eta}$ allows us to
sharply define the crossover behaviour in terms of a property of immediate physical significance, something that previous suggestions regarding a
crossover have not been able to provide~\cite{ref14,ref33}. We note that $T_{\eta}$ represents a useful estimate of the temperature
at which the Stokes-Einstein relation between viscosity and diffusion coefficients breaks down, as shown in Fig.~\ref{SE}.

\section{7. Conclusions}

In this paper we have demonstrated that the onset of viscous behaviour in simple liquids is a direct consequence of the
residual shear stress trapped in the individual local energy minima and its slow relaxation via activated transition
between the local minima. The variance of the inherent structure shear stress corresponds to the plateau height of the shear
stress autocorrelation function, a quantity more readily observable in experiments than the true infinite frequency modulus. We find that the variance of the inherent stress  quantity is independent of temperature and varies with density as a power law
$\rho^{\gamma}$ where the exponent $\gamma = 2(1+n/d)$ for d-dimensional liquids interacting by an inverse power law $r^{-n}$. We have established that the inherent structure stress corresponds to a long wavelength fluctuation, consistent with boundary
induced stress in the inherent structures. Such boundary effects are an inevitable consequence of the global rigidity that characterises any solid. Our results simply underscore the fact that all of the inherent structures, at the typically liquid state densities, are just such rigid solids. It is the fundamental duality of the viscous liquid state that, even as it exhibits all the fluid properties of a liquid, the slow stress relaxation is a direct consequence of the rigid solids that define the local minima of the liquid's energy landscape. The crossover in the mechanical linear response
of the liquid from the high temperature liquid to the viscous in which stress relaxation is dominated by the inherent structure
contribution, occurs at a high temperature (possibly above the melting point), similar in value to previously identified crossover temperatures~\cite{ref33,tarjus}.

Predicting the magnitude of $<\sigma^{2}_{xy}(IS)>$ or, more specifically, the factor $A$ in the expression $<\sigma_{xy}^{2}(IS)> = A\rho^{\gamma}$, remains a challenge. The absence of a temperature dependence in the distribution of the inherent stress indicates that the distribution is governed by the density of stress states. This is remarkable since the distribution of inherent energies does show strong temperature dependence which means that the same inherent stress distribution is being reproduced, at the different temperatures, by quite different selections of the inherent structures. The significance of the IS contribution to $<\sigma^{2}_{xy}(PL)>$ and, hence, to $G_{\infty}$ raise the question what is the relationship between the $G_{\infty}$ and the shear modulus of a glass, where the stress fluctuations associated with sampling many different IS would no longer contribute~\cite{wang,yoshino}. Understanding the the factors the determine the inherent stress  distribution, along with the magnitude of the stress change achievable by individual transitions between inherent structures and how the existence of residual stress influences our understanding of the shear modulus in the liquid and the glass represent an important set of open questions central to understanding the slow relaxation of shear stress in viscous liquids.

\noindent {\bf Acknowledgements}\\ It is a pleasure to acknowledge helpful discussions with Toby Hudson, Asaph Widmer-Cooper,
Mark Ediger and Cory O'Hern. This work was funded under the Discovery program of the Australian Research Council.
\\

\end{document}